\documentstyle[pra,aps,amsfonts,epsfig]{revtex}
\begin{document}
\draft
\twocolumn[\hsize\textwidth\columnwidth\hsize\csname 
@twocolumnfalse\endcsname
\title{Observation of Image Transfer and Phase Conjugation in Stimulated Down-Conversion} 
\author{P.H. Souto Ribeiro$^{*}$, D. P. Caetano and  M. P. Almeida}
\address{Instituto de F\'{\i}sica, Universidade Federal do Rio de 
Janeiro, Caixa Postal 68528, Rio de Janeiro, RJ 22945-970, Brazil}
\author{J. A. Huguenin, B. Coutinho dos Santos and A. Z. Khoury}
\address{Instituto de F\'{\i}sica,  Universidade Federal Fluminense, 
BR-24210-340 Niteroi, RJ, Brazil} 
\date{\today}
\maketitle
\begin{abstract}
We observe experimentally the transfer of angular spectrum and image formation
in the process of stimulated parametric down-conversion. Images and interference
patterns can be transferred either from the pump or the auxiliary laser beams to 
the stimulated down-converted one. The stimulated field propagates as the complex 
conjugate of the auxiliary laser. The phase conjugation is observed through intensity
pattern measurements.
\end{abstract}
]

The cavity-free stimulated down-conversion is a three-wave mixing
process(TWM). TWM have been viewed as a nonlinear optical process,
with promising applications for correcting wave front distortions.  
However, when this process was described by a quantum theory\cite{1}, a completely
different viewpoint was showed. The importance of the quantum approach is
closely related to the great success of the spontaneous parametric down-conversion
in Quantum Optics\cite{2} and the connection between the stimulated and the spontaneous processes.
The quantum properties of the stimulated emitted light has not yet been 
subject of intense investigation. Previous works show that it has coherence properties
of thermal and coherent states\cite{1,3}, but its individual quantum state was never
put forward.

In this work, we study
experimentally the transfer of angular spectrum and image formation in the
stimulated down-conversion, as predicted in Ref. \cite{4}. It was also predicted
that the stimulated field propagates as the complex conjugate of one of the
input fields. The understanding of the angular spectrum transfer has allowed 
the observation of the phase conjugation by direct intensity measurements.
Our results may have a classical counterpart (except for the spontaneous
emission part) in the framework of the TWM,
but they support the quantum theory of Ref.\cite{4}. The relation between
the angular spectrum of the pump, auxiliary(signal) and stimulated(idler) beams
were never demonstrated so far, to our best knowledge. These relations are
important because they will help us to separate the spontaneous from the stimulated
emission contributions in the total idler field. That will help us to understand
phase conjugation at the quantum level\cite{5}.

A laser pumps a non-linear crystal, producing twin photon pairs.
Another laser is aligned with one of the down-converted modes so that they overlap.
The second laser, which we will refer as to the auxiliary
laser, will stimulate the emission in those modes that best overlap, enhancing their intensities.
As a result the conjugate twin beam, which will be called the stimulated one, has its intensity 
enhanced and its spectral characteristics changed in the way it was discussed theoretically 
in Ref.\cite{4}.
The stimulated beam is generated from the superposition of the pump and auxiliary beams,
under the restrictions imposed by the phase matching on a particular non-linear medium.
Its field depends on the product of the pump and auxiliary laser fields. 
Its angular spectrum is given by the convolution between the angular spectrum
of the pump and the auxiliary lasers. 

According to Ref.\cite{4}, the intensity distribution of the stimulated (idler) 
beam is given by:
\begin{eqnarray}
\label{eq1}
I(\bbox{r}_{i}) \propto &&  \biggr\{ \int d\bbox{\rho} 
\left|{\cal W}_{p}(\bbox{\rho})\right|^{2} +
\\ \nonumber
&& \biggr| \int d\bbox{\rho}
{\cal W}_{p}(\bbox{\rho}) {\cal W}_{s}^{\star}(\bbox{\rho})
\exp \left[i |\bbox{\rho}_{i} - \bbox{\rho}|^{2}\frac{k_{i}}{2z}\right]
\biggr|^{2} \biggr\},
\end{eqnarray}
where $\bbox{r}_{i}$ = ($\bbox{\rho}_{i}$, $z$) is the position in the plane
transverse to the idler beam propagation at a distance $z$ from the crystal.
${\cal W}_{p}$ and ${\cal W}_{s}$ are respectively the pump and the auxiliary
lasers transverse field distributions at the crystal and k$_{i}$ is the idler
wavenumber.

In our experiment, the spontaneous emission is much smaller than the stimulated
one so that the first term in Eq.\ref{eq1} can be neglected. We are going to analyze
two particular situations. In the first one, the auxiliary laser amplitude distribution
at the crystal can be considered constant. Therefore, the resulting idler intensity
is given by:
\begin{eqnarray}
\label{eq2}
I(\bbox{r}_{i}) \propto \biggr| \int d\bbox{\rho}
{\cal W}_{p}(\bbox{\rho}) 
\exp \left[i |\bbox{\rho}_{i} - \bbox{\rho}|^{2}\frac{k_{i}}{2z}\right]
\biggr|^{2} ,
\end{eqnarray}

This intensity distribution is the same as the pump's after it has propagated to
a plane situated at a distance $z$ from the crystal. The only difference is that
the propagation goes with the idler wavevector. 

In the second situation, the pump beam amplitude distribution at the crystal can be
approximated by a constant. In this case, the idler intensity is given by:

\begin{eqnarray}
\label{eq3}
I(\bbox{r}_{i}) \propto \biggr| \int d\bbox{\rho}
{\cal W}_{s}^{\star}(\bbox{\rho}) 
\exp \left[i |\bbox{\rho}_{i} - \bbox{\rho}|^{2}\frac{k_{i}}{2z}\right]
\biggr|^{2} ,
\end{eqnarray}
In analogy to the previous case, the intensity distribution of the idler should follow 
the auxiliary laser one after propagation to a plane situated at a distance $z$ from the crystal.
However, now it depends on the complex conjugate of the auxiliary laser field amplitude at the crystal.
The above equations were derived within the Fresnel approximation. In this regime, the phase
conjugation does not affect the transferred image in a simple way and it will resemble that on the 
auxiliary laser. However, if the Fraunhofer limit can be taken, the image will be inversed, as we shall
see. In the Fraunhofer limit the phases can be written as:
\begin{eqnarray}
\label{eq4}
\exp \left[i |\bbox{\rho}_{i} - \bbox{\rho}|^{2}\frac{k_{i}}{2z}\right]
\rightarrow 
\exp\left(i \bbox{\rho}_{i} \cdot \bbox{\rho} \frac{k_{i}}{z}\right).
\end{eqnarray}
and Eq.\ref{eq3} turns into
\begin{eqnarray}
\label{eq5}
I(\bbox{r}_{i}) \propto \biggr| \int d\bbox{\rho}
{\cal W}_{s}^{\star}(\bbox{\rho}) 
\exp\left(i \bbox{\rho}_{i} \cdot \bbox{\rho} \frac{k_{i}}{z}\right)
\biggr|^{2} ,
\end{eqnarray}

Taking the complex conjugate of the argument of the square modulus, the result
remains unchanged:

\begin{eqnarray}
\label{eq6}
I(\bbox{r}_{i}) \propto \biggr| \int d\bbox{\rho}
{\cal W}_{s}(\bbox{\rho}) 
\exp\left(-i \bbox{\rho}_{i} \cdot \bbox{\rho} \frac{k_{i}}{z}\right)
\biggr|^{2} .
\end{eqnarray}

It is easy to see that the intensity profile of the idler beam will have the same shape
as the auxiliary laser one,
but with $\bbox{\rho}_{i}$ replaced by $-\bbox{\rho}_{i}$. That is to say,
the image on the idler beam is inverted with respect to that on the auxiliary laser beam.

\begin{figure}[h]
\vspace*{3.3cm}
\hspace*{1cm}
\epsfig{file=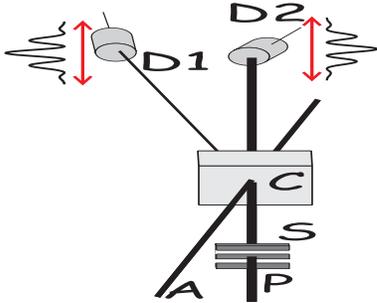,width=5cm,height=4cm}
\caption{Sketch of the experiment. P is the pump beam, A is the auxiliary beam S is the double-slit, C is the nonlinear crystal, D1 and D2 are detectors.}
\label{fig1a}
\end{figure}

\begin{figure}[h]
\vspace*{3.5cm}
\hspace*{1cm}
\epsfig{file=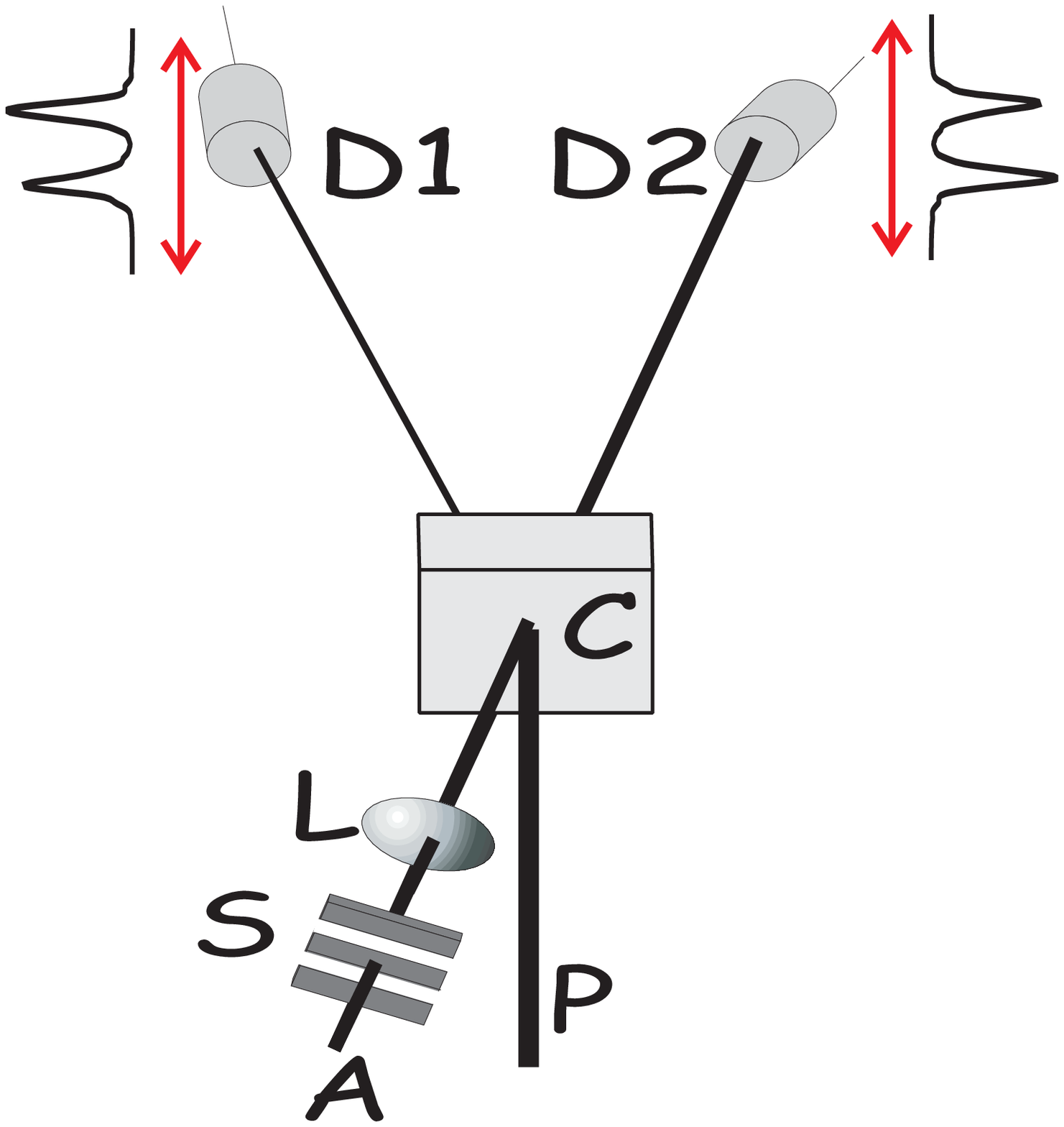,width=5cm,height=4cm}
\caption{Sketch of the experiment. P is the pump beam, A is the auxiliary beam S is the double-slit, L is the lens, C is the nonlinear crystal, D1 and D2 are detectors.}
\label{fig1b}
\end{figure}

A sketch of the experimental set-up is shown in Figs.\ref{fig1a} and Fig.\ref{fig1b}. 
The vertically polarized He-Cd laser pumps a BBO crystal. 
Spontaneous parametric down-conversion
takes place and twin photons converted from the pumping wavelength 442nm, to
the signal 845nm and idler 925nm are produced. Because of the type II
phase matching, the signal beam is vertically polarized and the idler is horizontally
polarized. We utilize photon counting modules (EG\& GSPCM-AQ151) for detecting signal and
idler photons in coincidence. In the signal side of the detection, a 10nm bandwidth interference
filter centered in 845nm is utilized, while in the idler side the bandwidth is 50nm centered in 
920nm. This is the first step for obtaining stimulated down-conversion. 

The second step is to carefully align an auxiliary diode laser with the 845nm signal beam.
The diode laser wavelength can be tuned from 840 to about 850nm, by temperature and
current control.
When the alignment is good enough, the intensity of the idler is immediately multiplied
by a large factor, indicating the amplification of the emission in this particular mode.
The factor of amplification is determined by the coupling between the auxiliary and
signal down-conversion modes. In our set-up it was about 300.

\begin{figure}[h]
\vspace*{2.5cm}
\epsfig{file=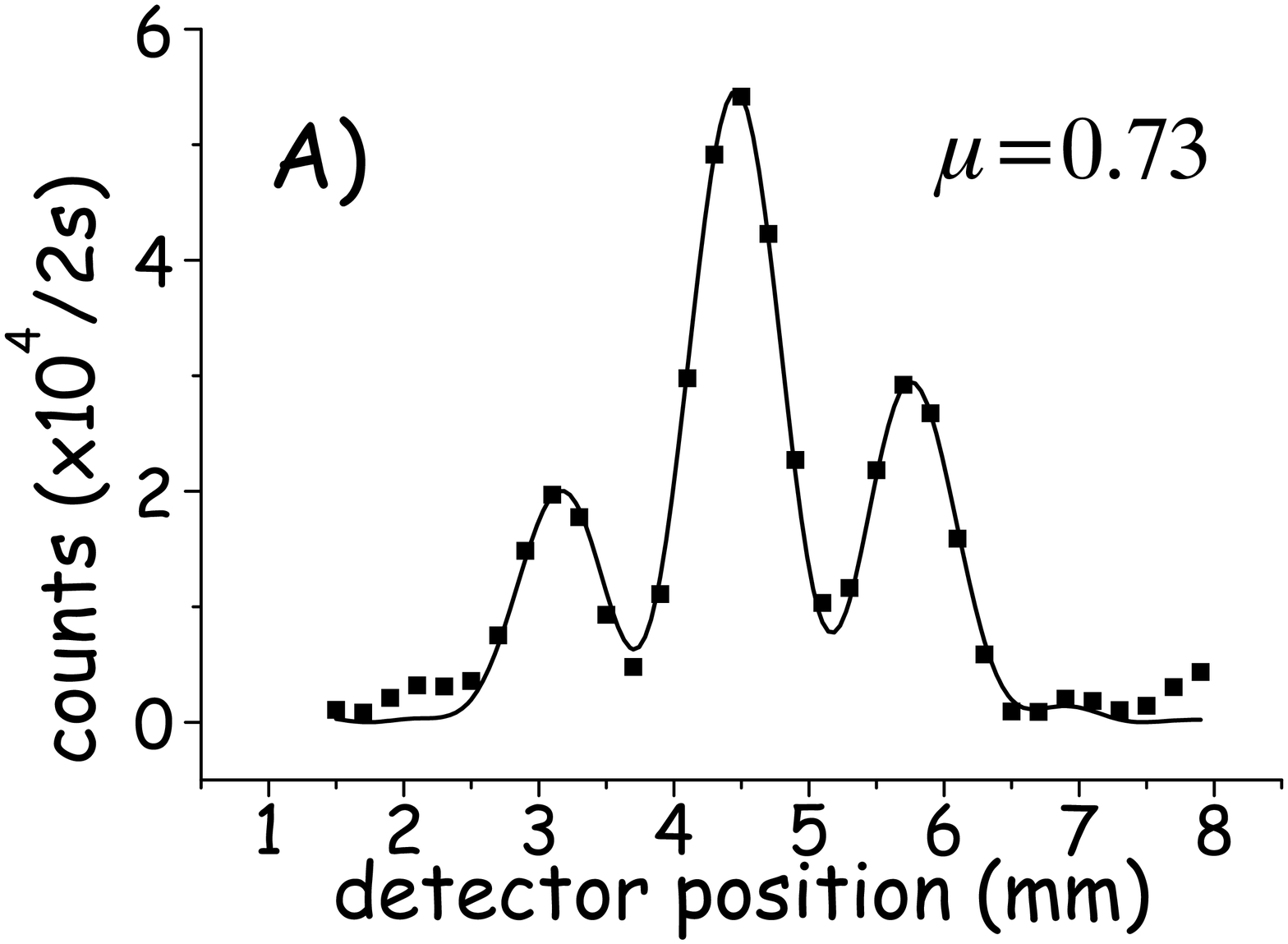,width=7cm,height=3cm}
\vspace*{3cm}
\epsfig{file=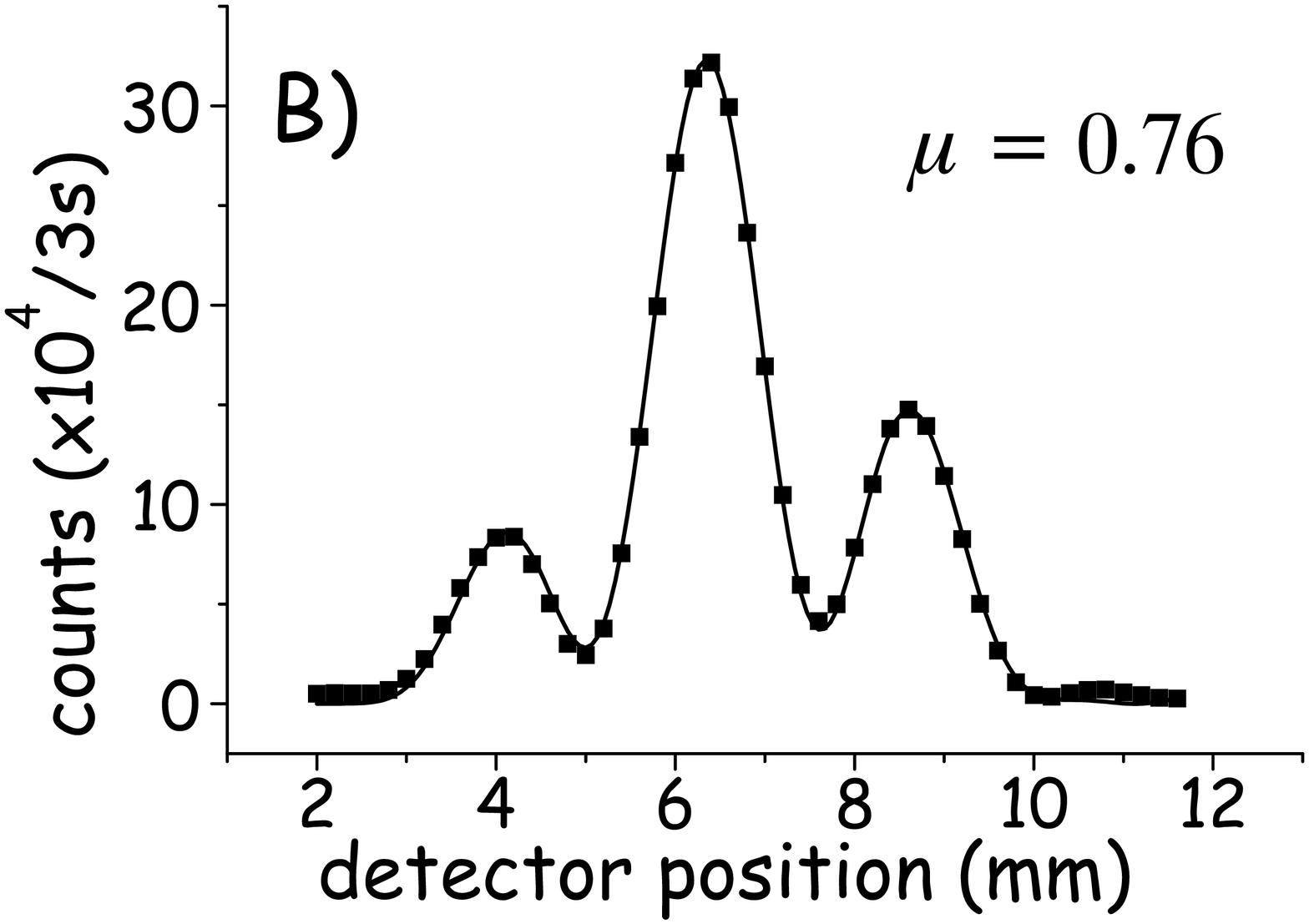,width=7cm,height=3cm}
\caption{Vertical intensity profile for A) pump beam with a double-slit 
inserted 15cm before the crystal and B) stimulated idler beam. Solid lines are
fittings to typical double-slits pattern functions.}
\label{fig2}
\end{figure}

The experiment consists of placing masks on the pump(or auxiliary) beam before the
crystal and observing the relationship between pump(or auxiliary) and idler intensity
distributions on a plane situated at the same distance from the crystal for both beams.
First, a double-slit is placed on the pump beam, about 15 cm before the crystal, so that its
spectrum carries the information of the passage through the slits during the non-linear interaction.
See Fig.\ref{fig1a}.
After the crystal the double-slit interference pattern is formed on the pump beam. 
We have measured this
pattern on a plane situated about 80cm from the crystal. This is shown in Fig. \ref{fig2}A.
According to Eq. \ref{eq2}, if the field amplitude of the auxiliary laser beam can 
be approximated by a constant, the interference pattern of the pump beam is
transferred to the stimulated idler beam. Note that, while the shape of the intensity
distribution of the idler should {\em perfectly} (in the ideal case) follow that
of the pump, the propagation of the field from the crystal to detection depends on the 
idler wavelength. As a result, the pattern is about two times larger than that of the pump.
We have measured it also on a plane placed about
80cm from the crystal. It is shown in Fig. \ref{fig2}B. 

\begin{figure}[h]
\vspace*{2.5cm}
\epsfig{file=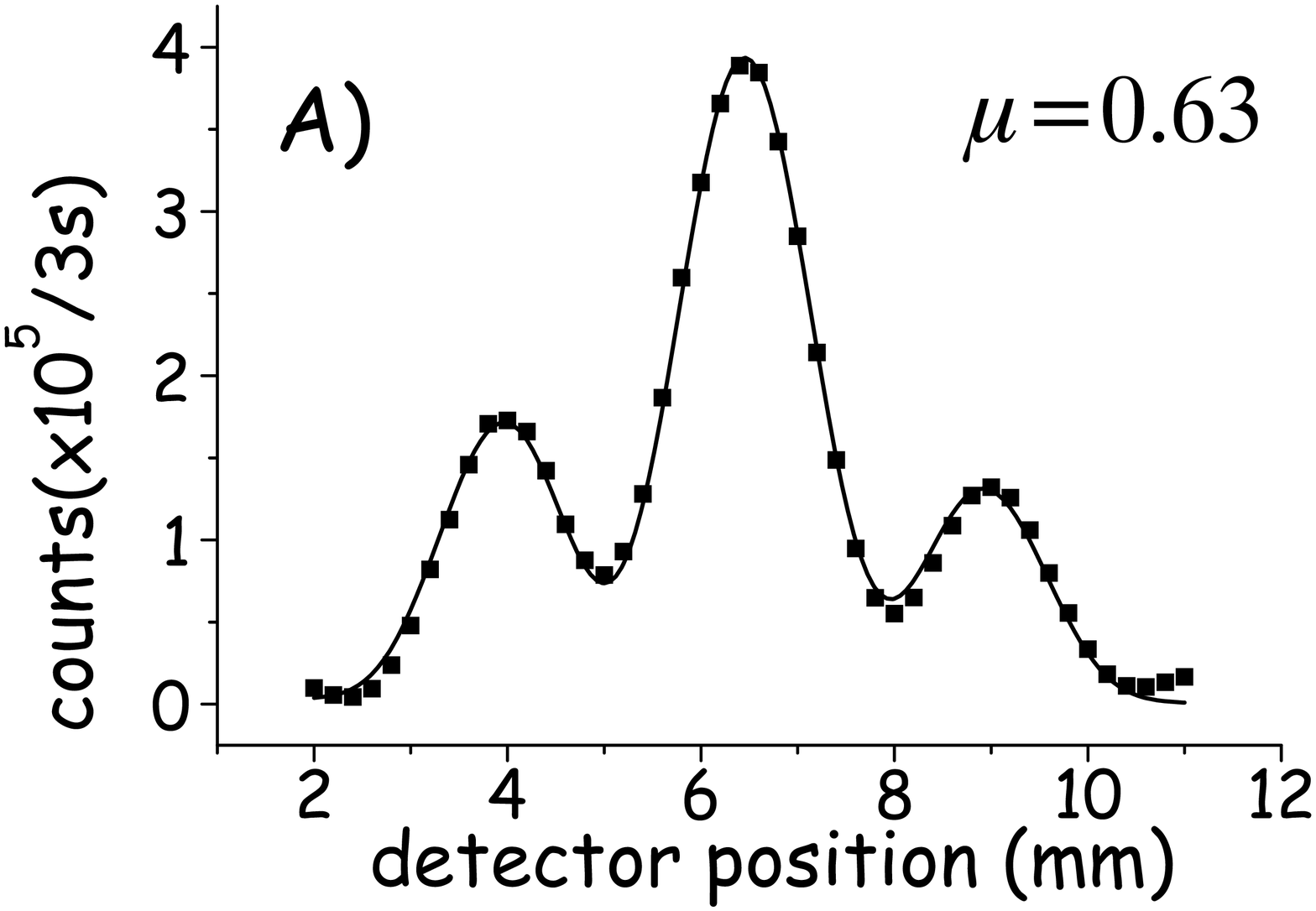,width=7cm,height=3cm}
\vspace*{3cm}
\epsfig{file=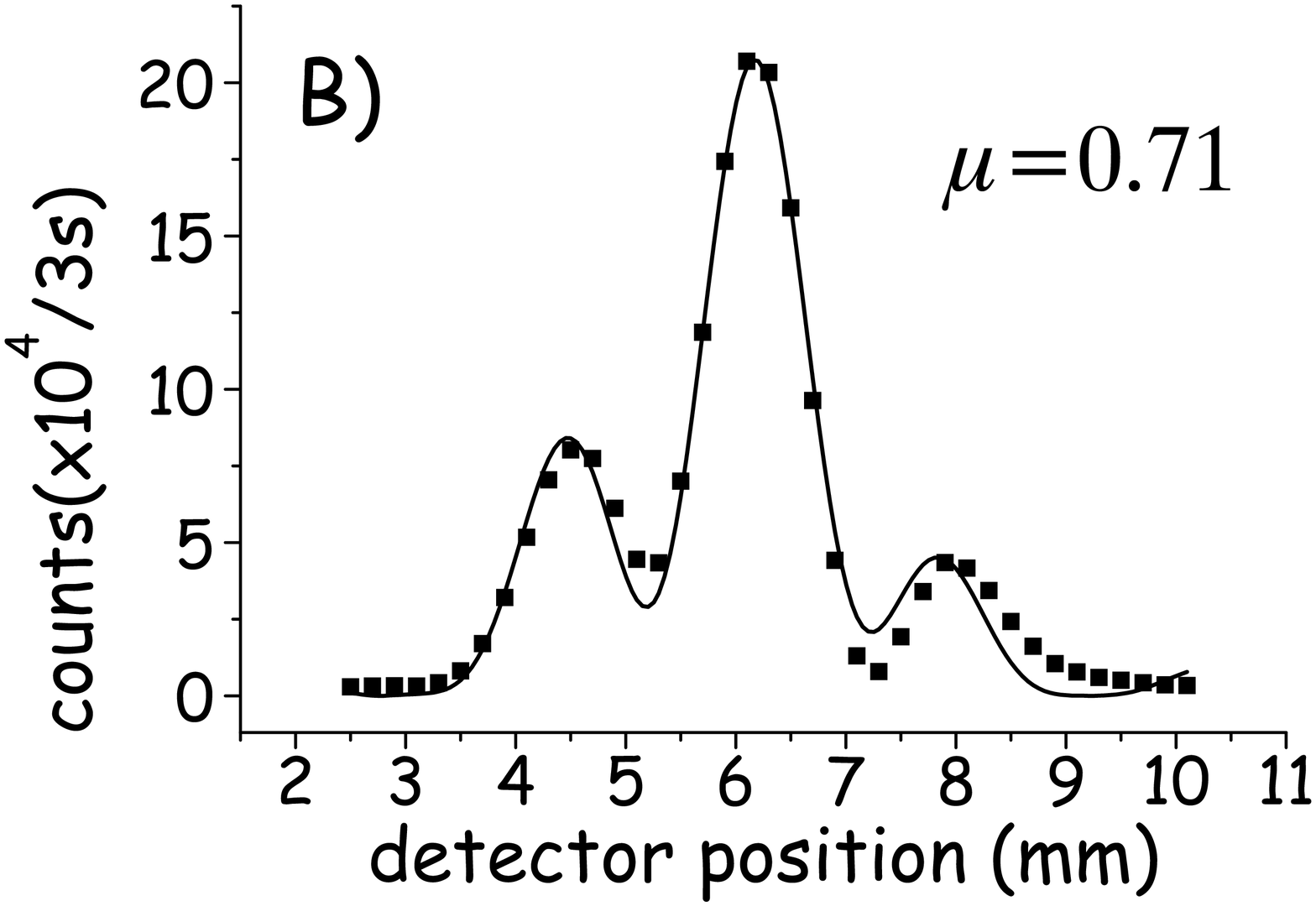,width=7cm,height=3cm}
\caption{Vertical intensity profile A) auxiliary beam with a double-slit
inserted  15cm before the crystal and B) stimulated idler beam. Solid lines are
fittings to typical double-slits pattern functions.}
\label{fig4}
\end{figure}

In a second set of measurements, we have placed the double-slit in the path of the
auxiliary laser 15cm before the crystal. The pattern formed on the auxiliary beam
80cm after the crystal is shown in Fig. \ref{fig4}A. 
Once again, the intensity pattern is transferred to the
stimulated idler beam, according to Eq. \ref{eq3} for a constant pump field
amplitude distribution. The measured pattern for the idler beam on a plane 80cm far from the crystal
is shown in Fig. \ref{fig4}B. The pattern follows that of the auxiliary beam without inversion
of the image, because this is the Fresnel diffraction.
These results show that the angular spectrum from the pump
and the auxiliary lasers are transferred to the stimulated down-conversion. Note that
as the double-slit is small(about 0.4mm width and separated by 0.2mm)and it is placed 15cm 
before the crystal, the field inside the crystal is not an image of the slits with two bright spots that
illuminates the crystal during the interaction,  but rather a more complicated spatial
distribution.

\begin{figure}[h]
\vspace*{2.5cm}
\epsfig{file=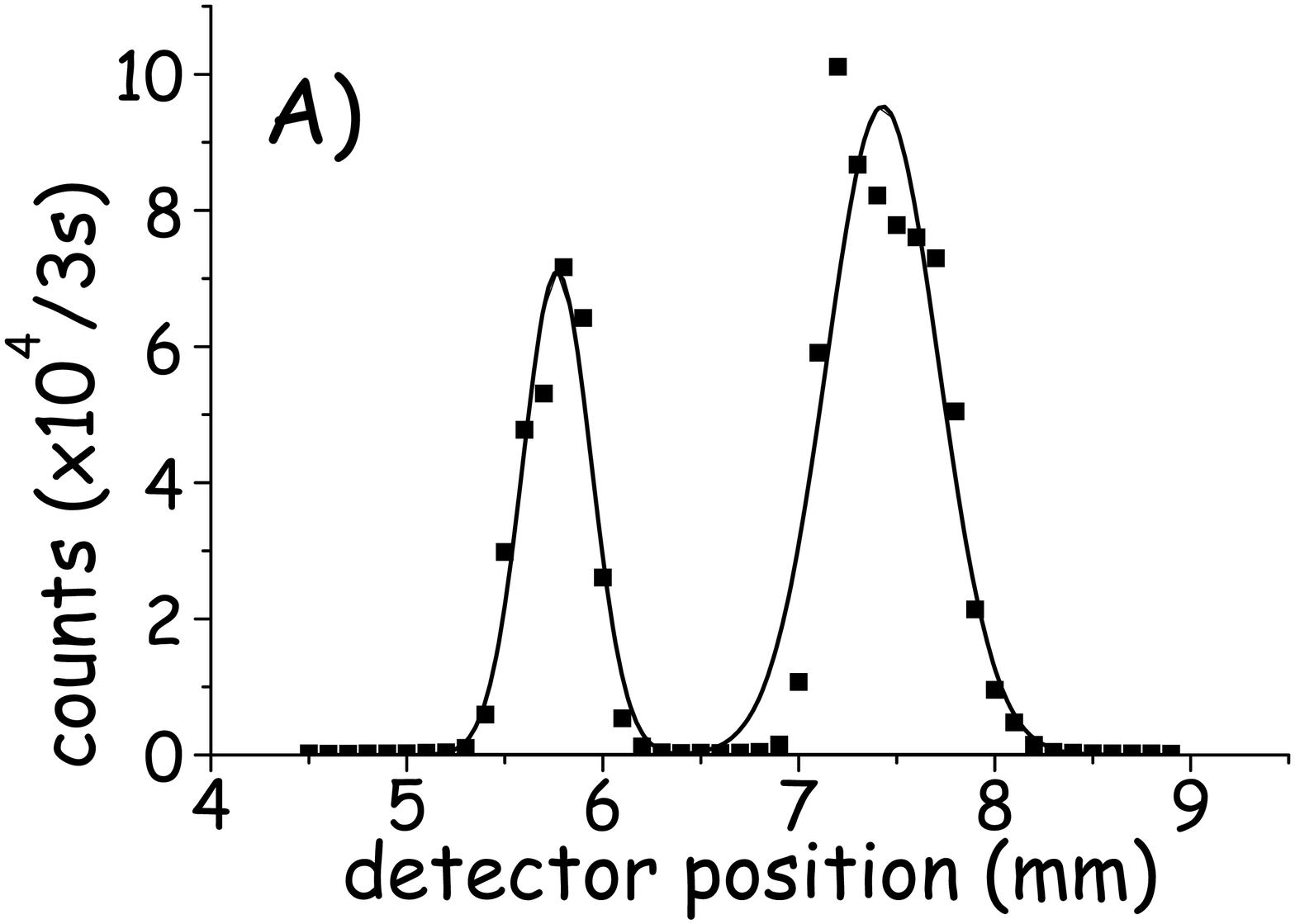,width=7cm,height=3cm}
\vspace*{3cm}
\epsfig{file=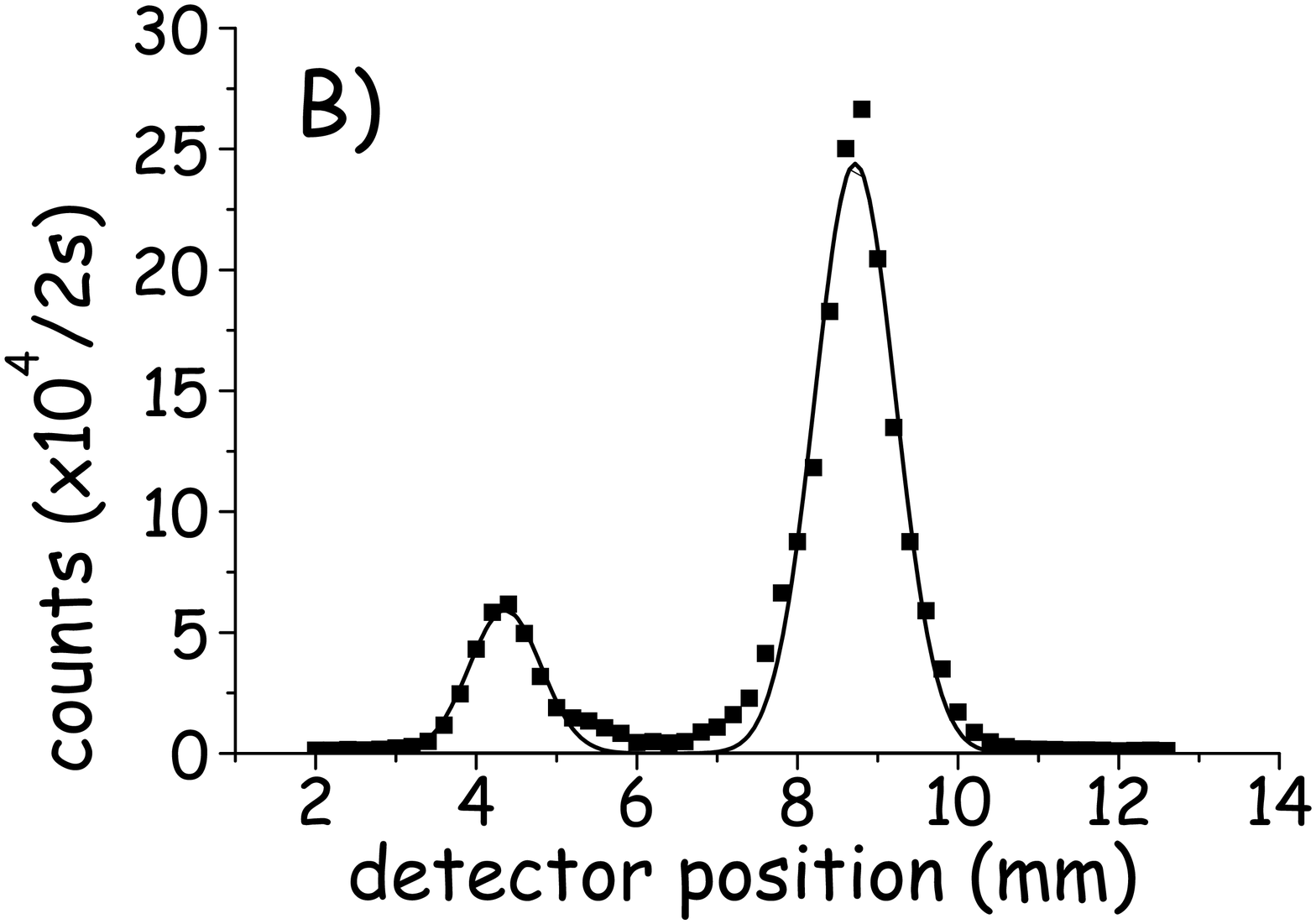,width=7cm,height=3cm}
\caption{Vertical intensity profile for A) pump beam when a double-slit image is
focused in the detection plane 80cm after the crystal B) stimulated idler beam.
Solid lines are only guides for the eye.}
\label{fig6}
\end{figure}

If the angular spectrum is transferred from the pump and auxiliary lasers
to the stimulated idler beam, images rather than interference patterns observed 
on the intensity profile of those
laser beams should also be transferred to the stimulated beam. 
In order to test it, we have projected the double-slit image on the detection plane,
when it was inserted both on the pump and the auxiliary beams. See Fig.\ref{fig1b},
where slits and the lens were inserted on the auxiliary beam. This is the simplest
one-dimensional image we could utilize.
The double-slits are now larger than before. For the image pattern we have utilized
slits of about 1mm width separated 1 mm from each other. The pump intensity distribution in the
vertical direction at the image plane is shown in Fig. \ref{fig6}A, when the slits and the lens
are inserted on the pump beam.
In Fig. \ref{fig6}B the transferred image can be observed at the idler stimulated intensity
profile. Like before, the image formed on the idler beam is larger than
that on the pump beam because of the influence of the wavelength on the propagation.
The intensity peaks were made asymmetric so that we can see how the image is transferred
from the pump and auxiliary lasers to the idler beam.

\begin{figure}[h]
\vspace*{2.5cm}
\epsfig{file=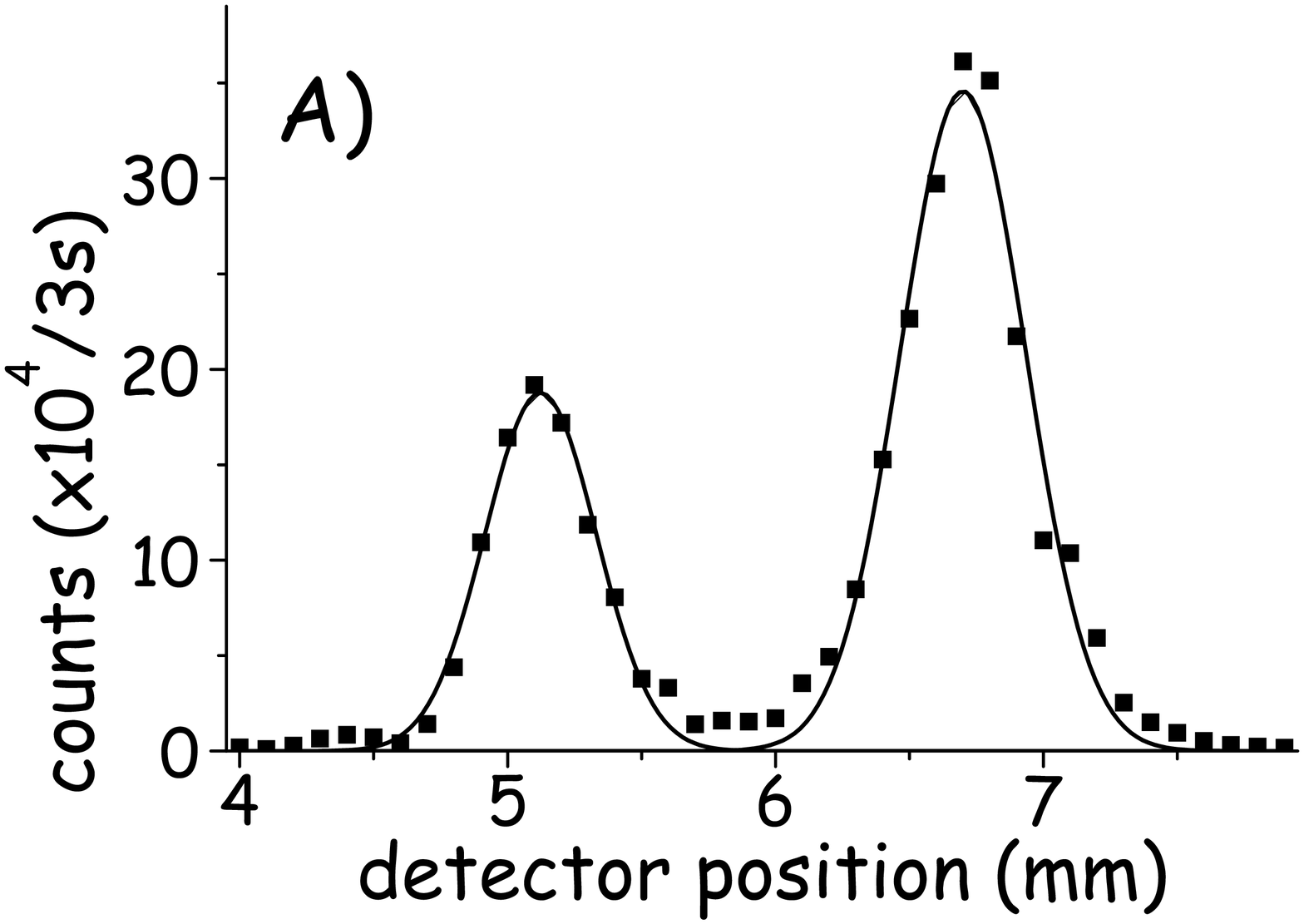,width=7cm,height=3cm}
\vspace*{3cm}
\epsfig{file=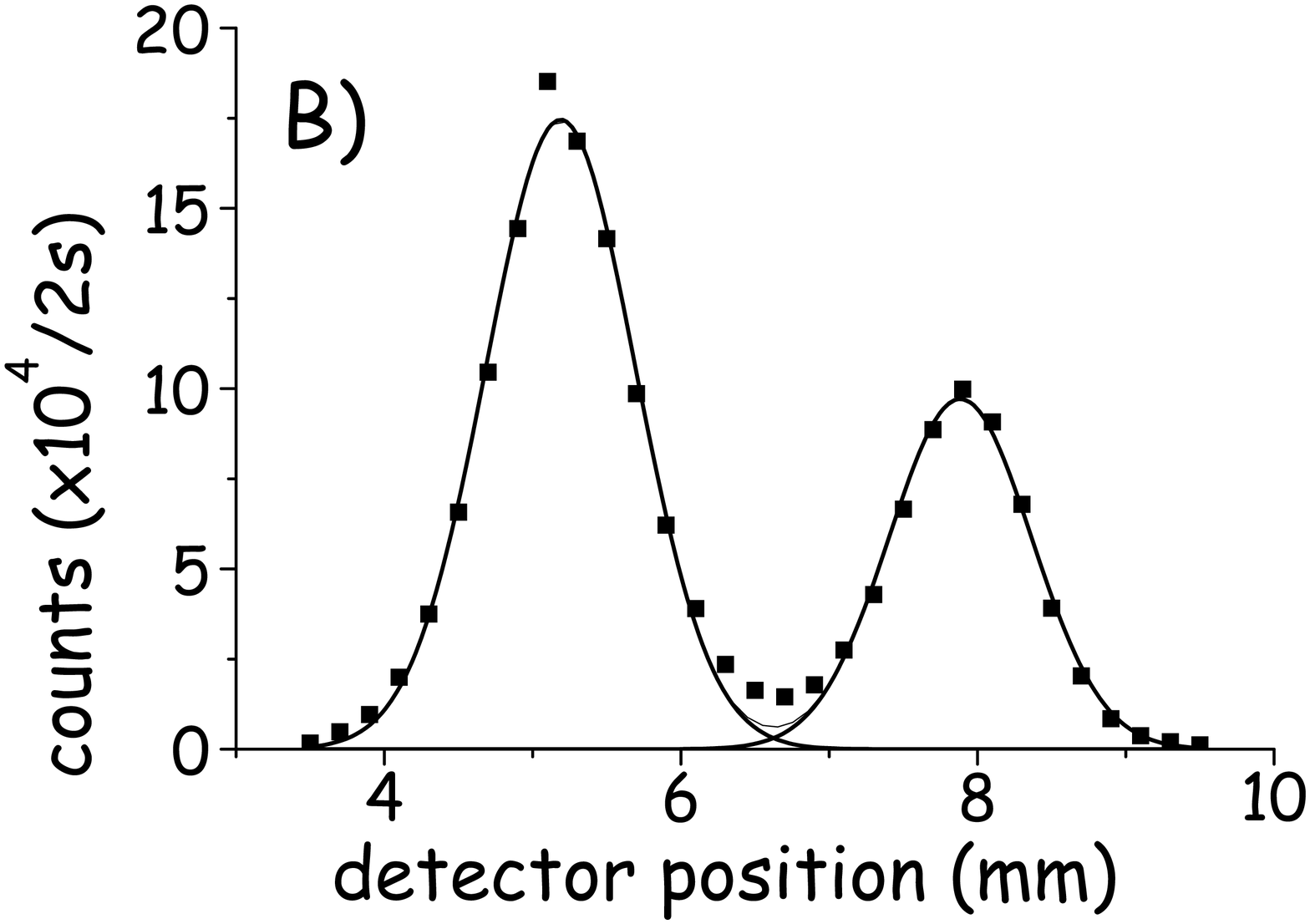,width=7cm,height=3cm}
\caption{Vertical intensity profile for A) auxiliary beam when a double-slit image is
focused in the detection plane 80cm after the crystal and B) stimulated idler beam.
Solid lines are only guides for the eye.}
\label{fig8}
\end{figure}

The same procedure was repeated placing the slits on the auxiliary beam and obtaining
its image after the crystal, through focalization by a lens . Its intensity distribution
is shown in Fig. \ref{fig8}A. However, the image transferred to the idler stimulated beam
does not exactly repeat the profile of the auxiliary beam. As the intensity of the peaks
are different in Fig.\ref{fig8}A, it is possible to see that the transferred image is inverted
in Fig.\ref{fig8}B. This is a consequence of the phase conjugation as predicted by Eq.\ref{eq6}. 
Now, the focalization of the auxiliary beam inside the crystal has taken us to the Fraunhofer
limit because its larger dimension($\bbox{\rho}_{max}\sim 0.1 mm$) is much
smaller than the distance between the crystal and the detection plane($z\sim 1m$) 
$\bbox{\rho}_{max}^{2} \ll \lambda z$, with $\lambda \sim 10^{-6}m$.

\begin{figure}[h]
\vspace*{2.5cm}
\epsfig{file=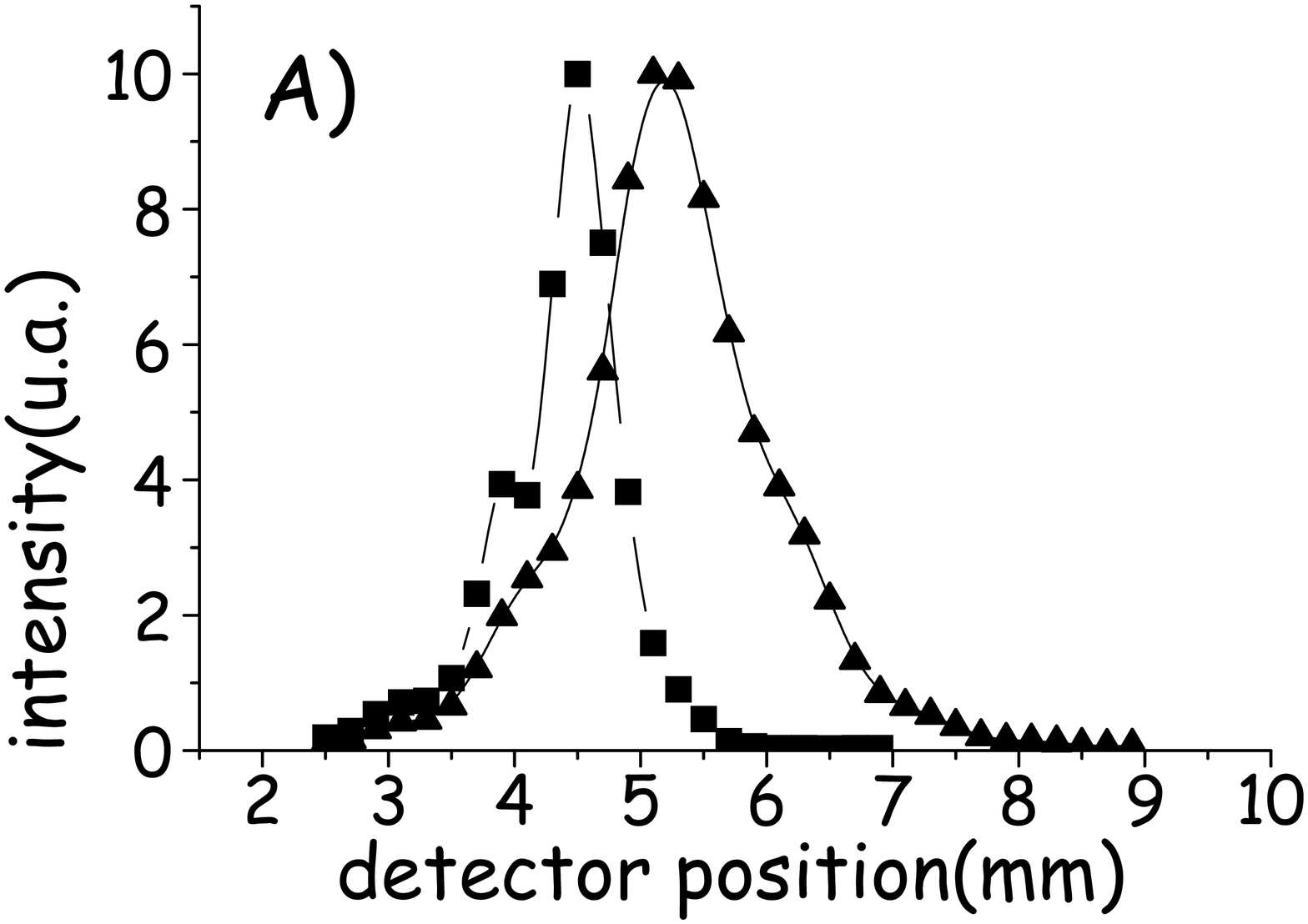,width=7cm,height=3cm}
\vspace*{3cm}
\epsfig{file=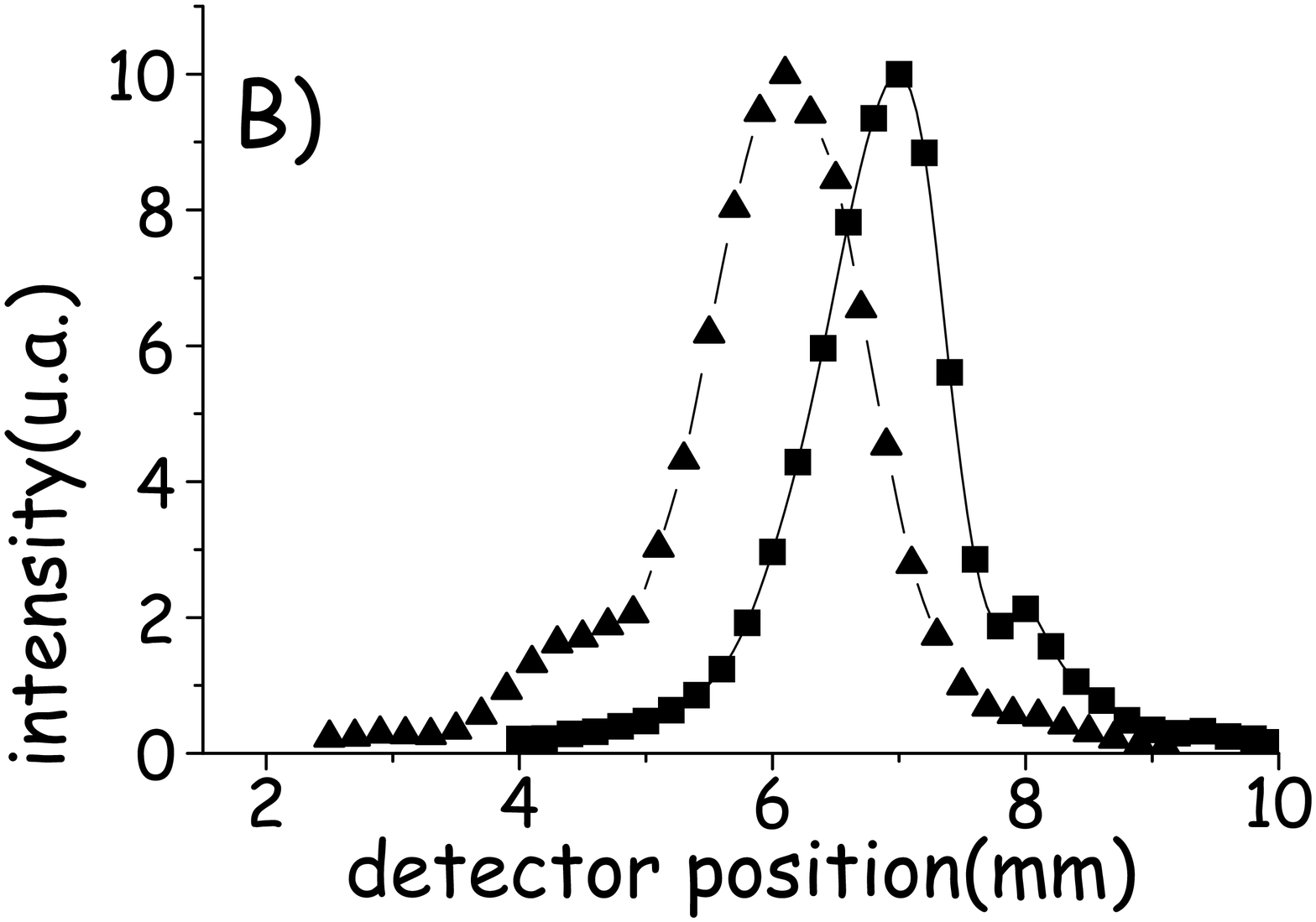,width=7cm,height=3cm}
\caption{Vertical intensity profile for A) auxiliary beam with(squares) and without(triangles)
an inserted blade before the crystal and B) the stimulated idler beam with(squares) and 
without(triangles) the blade in the auxiliary.
Solid lines are only guides for the eye.}
\label{fig10}
\end{figure}

We have performed other measurements, for supporting the predictions of Eq.\ref{eq6}.
A sharp blade was 
placed on the auxiliary beam before the crystal, so that about a half of the beam was covered 
by the blade. Using a lens, the image of the half beam was projected on the detection plane, 
placed 80cm from the crystal. The intensity distributions of the auxiliary and idler beams were
measured with and without the presence of the blade. This is shown in Figs. \ref{fig10}A and 
\ref{fig10}B for the auxiliary and idler beams respectively. The effect of the phase conjugation
is then confirmed by the displacement of the intensity peaks in opposite senses. In other
words, the dark half of the beam, due to the presence of the blade, has appeared in
opposite sides for the image formed on the auxiliary beam and the transferred image
formed by the conjugate field on the idler beam.

The results presented above can be understood in terms of the Eq. \ref{eq1}.
The equation says that the intensity profile of the idler stimulated down-conversion
is given by the product of the pump and auxiliary lasers amplitude distributions at the
crystal, propagated by a Fresnel propagator to the observation plane after the crystal.
The propagation is performed through the idler wavevector as it should be, but when the
pump(auxiliary) field amplitude at the crystal is constant the transverse intensity distribution
of the idler will follow that of the auxiliary(pump) after propagation to the same observation plane.

We would like to emphasize two points. First, the idler intensity depends on the complex conjugate
of the auxiliary field. Even though the dependence comes on the
intensity, so that the square modulus of the auxiliary field is calculated, the phase conjugation has
observable effects in the idler intensity. 
This is because the conjugation changes the propagation of the field. 
In the above reported case, it has been responsible for the inversion of the idler image
compared to the auxiliary laser one, as it was predicted by Eq.\ref{eq6}. However, this equation
is only valid in the Fraunhofer limit, far enough from the source(crystal). In the Fresnel limit
the image is not inverted because the propagation depends on 
$\exp \left[i |\bbox{\rho}_{i} - \bbox{\rho}|^{2}\frac{k_{i}}{2z}\right]$.

The second point is that the theoretical treatment adopted is fully quantum mechanical,
In the present work, we are not exploring the quantum aspects of the process, but it can
be used to study the quantum aspects of the image formation and phase conjugation in the
stimulated parametric down-conversion. In this case we should probably use weaker auxiliary
beams and correlation detection schemes.

In conclusion, we have observed experimentally the transfer of the angular spectrum from the pump
and auxiliary lasers to the idler beam in the process of cavity-free stimulated
down-conversion. The transfer of the angular spectrum is observed through the 
transfer of images and interference patterns. The experimental results are in
agreement with the theoretical predictions of Ref. \cite{4}.

We have also demonstrated experimentally that the idler field propagates as the complex 
conjugate of the auxiliary laser field, even though they are completely different modes
with different wavelengths and polarizations. The demonstrated properties can lead to
applications where  images coded by a random medium (key) may be transferred from the 
auxiliary beam to the idler. The decoding may be achieved by propagation of the idler through
the key, for example. They have also shown to be useful for studying the process of phase 
conjugation at the quantum level.

Financial support was provided by Brazilian 
agencies CNPq, PRONEX, FAPERJ and FUJB.

\end{document}